\documentclass[12pt,a4paper]{article}
\usepackage{amsmath}         
\usepackage{amssymb}
\usepackage{bbm}             
\usepackage{amsthm}
\usepackage{graphicx}
\usepackage{hyperref}       
\usepackage{soul,color}

\newcommand{\ist}[1]{\overset{\footnotesize(\ref{#1})}{=}}
\newcommand{\iist}[2]{\overset{^{(\ref{#1})}}{\underset{^{(\ref{#2})}}{=}}}
\newcommand{\gehtzu}[1]{\stackrel{\footnotesize{#1}}{\to}}

\begin{document}

\title{From soft Dirac monopoles to the Dirac equation}
\author{Manfried Faber,\\ Technische Universit\"at Wien, Atominstitut\\Stadionallee 2, 1020 Wien, Austria}

\maketitle

\begin{abstract}
In the model of topological particles we have four types of topologically stable dual Dirac monopoles with soft core and finite mass. We discuss the steps how to get a Dirac equation for these particles. We show for the free and the interacting case that we arrive at the Dirac equation in the limit, where the soft solitons approach singular dual Dirac monopoles.
\end{abstract}

\section{Introduction}
According to Maxwell's electrodynamics the electric self-energy of point-like electrons is infinite. Schrödinger was originally~\cite{Schrodinger:1926aaa} inclined to a standing wave interpretation of electrons in his equation, whereas Bohr, Heisenberg, Pauli and Born~\cite{PEROVIC2006275} insisted in the corpuscular nature of electrons, where electrons have to be seen as point-like objects and the square of the wave function describes the probability of its position only.

Schrödinger and Dirac equations describe very successfully the electromagnetic interaction between electron and proton without taking explicitly into account the electromagnetic self-energy of point-like or extended electrons. This self-energy should exist according to Maxwell. It is absorbed in the observable mass $m_e$  of electrons at rest. Despite Schrödinger and Dirac equations do not treat the electromagnetic field self consistently, neglecting the $F_{\mu\nu}F^{\mu\nu}$ contribution, they predict the properties of the hydrogen atom very well and even shaped our understanding of atomic and condensed matter physics. Only the tiny Lamb shift is not described in the Dirac equation.

The Lagrangian of Quantum Electrodynamics (QED) takes the energy density of the electromagnetic field into account. The divergences of point-like charges are cured, as Kramers suggested 1947 at the Shelter Island Conference, by subtracting compensating infinities from the naive QED Lagrangian. Finally, the theory gets consistent only by respecting the running of the couplings, a feature which is in excellent agreement with high energy scattering experiments.

On the other hand, for a classical relativistic model with stable extended solitons of finite mass and quantised charges, as discussed in~\cite{Faber:1999ia,Faber:2022zwv}, there is no need to absorb the self-energy of elementary charges in a mass term adjusted to the experimental electron mass. Charges and their fields are described by the same degrees of freedom and distinguished by topological quantum numbers only. Therefore the Hamiltonian contains the self-energy of charges, the energy of the electromagnetic field and the interaction energy of charges and electromagnetic fields.

Up to now MTP was formulated as a classical model. Thus, there is urgent need to harmonise it with quantum physics. A classical model misses the quantum properties, well described by the Schr\"odinger and the Dirac equation. Quantum effects for extended solitons can as well be formulated by a path integral. All these descriptions require the electric interaction energy. This can be directly determined within the soliton model. Due to the finite size of solitons we expect at distances of several femtometers deviations from the potential between point-like charges. Inserting these potential modifications into the Schrödinger or Dirac equation should lead to a shift of energy levels, especially for s-states. It will be very interesting to compare these shifts with the experimental spectra and especially the Lamb shift.

There are essential differences of the soliton formulation of hydrogen and positronium to the conventional Schr\"odinger equation. Here we have to especially point out the differences in the field degrees of freedom (dofs). The soliton model describes charges and electromagnetic fields by the same dofs, by an SO(3)-valued scalar field, in strong distinction from the conventional description with Grassmann-valued $\psi$-fields and gauge fields $A_\mu$. Moreover the vector fields of the soliton model describe dual photons. The attempt to describe essential properties of nature by our soliton model, implies immediately the question, how one can understand within this picture the impressive success of the conventional determination of the properties of atomic systems with the Schr\"odinger and Dirac equation. Therefore, we try to analyse in this article, by which steps we can arrive at the conventional Dirac equation starting from our soliton model. Moreover, this investigation sheds light onto the above suggested insertion of the soliton-antisoliton potential energy into the Schrödinger and Dirac equation.

In Sect.~2 we present a very short formulation of the soliton model and some of its properties. Extended versions of this model were published in Refs.~\cite{Faber:1999ia,Faber:2022zwv}. The attempt to get the conventional Dirac equation we realise in two steps. First we concentrate on the free Dirac equation in Sect.~\ref{SectFree} using the relativistic properties of solitons. Then we introduce the interaction with an external field by artificially separating the dofs of charged solitons and their fields in Sect.~\ref{SectInteract}. Since the conventional Dirac equation is using a pure $1/r$-potential, it is sufficient to formulate this separation in the limit of point-like charges, where this separation can be formulated more easily. Due to charge quantisation the generalisation to extended charges is obvious if the distance between solitons is large compared to their size. Finally, we discuss the question, why the conventional Dirac equation can be formulated with gauge fields $A_\mu$, despite the fact that the soliton model is formulated with dual vector fields.

\section{Model of Topological Particles}
The Model of Topological Particles (MTP)~\cite{Faber:1999ia,Faber:2022zwv} formulates electric mono\-poles and their fields as topologically stable solitons of finite mass and is characterised by two topological quantum numbers which can be related to spin and charge. There exist four classes of such solitons without any singularities with the quantum numbers of the components of Dirac fermions.

For the description of electromagnetic phenomena and especially of solitons in 3+1D with long-range coulombic interaction the MTP uses the three degrees of freedom of an SO(3)-field only. But the calculations are simpler using the $2\times2$-matrices
\begin{equation}\label{Grundfeld}
Q(x)=q_0(x)-\mathrm i\vec\sigma\vec q(x)\in~\textrm{SU(2)},\quad
q_0^2+\vec q^2=1,\end{equation}
matrices in the double covering group of SO(3). In the model of topological particles (MTP), see Ref.~\cite{Faber:2022zwv}, the Lagrangian reads
\begin{equation}\label{Lagr4D}\hspace{-4mm}
\mathcal L_\mathrm{MTP}:=-\frac{\alpha_f\hbar c_0}{4\pi}\left(\frac{1}{4}\,\vec R_{\mu\nu}
\vec R^{\mu\nu}+\Lambda(q_0)\right),\quad\Lambda(q_0):=\frac{q_0^6}{r_0^4},\quad
\alpha_f:=\frac{e_0^2}{4\pi\varepsilon_0\hbar c_0}\end{equation}
where the curvature field $\vec R_{\mu\nu}$,
\begin{equation}\label{RSU2}
\vec R_{\mu\nu}:=\vec\Gamma_\mu\times\vec\Gamma_\nu
\end{equation}
is the area density on \textrm{SU(2)}, the Jacobian for the map of areas on M$^4$ to areas on S$^3$. It is algebra valued as indicated by the vector symbol. The vector field~\footnote{For simplicity we omit the dot for scalar products in three dimensions, if it does not lead to ambiguities. The $\times$ symbol acts always in three dimensional algebra and not in coordinate space.}
\begin{equation}\label{GammaSU2}
\partial_\mu Q(x)=:-\mathrm i\vec\Gamma_\mu(x)\,\vec\sigma Q(x)
\end{equation}
is the affine connection in the tangential space of SU(2).

With the generalised velocity $\vec\Gamma_\mu$ we get the energy momentum tensor
\begin{equation}\begin{aligned}\label{DefEMT}
{\Theta^\mu}_\nu(x)&:=\frac{\partial\mathcal L(x)}{\partial\vec\Gamma_\mu}\,
\vec\Gamma_\nu-\mathcal L(x)\,\delta^\mu_\nu=\\
&\iist{Lagr4D}{RSU2}-\frac{\alpha_f\hbar c_0}{4\pi}
\left\{\left(\vec\Gamma_\nu\times\vec\Gamma_\sigma\right)
\left(\vec\Gamma^\mu\times\vec\Gamma^\sigma\right)\right\}-
\mathcal L(x)\,\delta^\mu_\nu .
\end{aligned}\end{equation}
which is automatically symmetric.

This model is a generalisation of the Sine-Gordon model \cite{remoissenet:2003} from 1+1D and 1 dof to 3+1D and 3 dofs. It can be considered as a modification of the Skyrme model~\cite{skyrme:1958vn,Adam:2013tga,Gudnason:2022jkn} describing solitons with long-range forces in 3+1D space-time with Minkowski metric. Further, it can be seen as a model for soft dual Dirac monopoles~\cite{dirac:1948um,wu:1975vq}, this means Dirac monopoles without any singularities, without Dirac string~\cite{wu:1975vq,wu:1976qk} and without singularity in the center. Another model with monopoles without singularities is the Georgi-Glashow model~\cite{Bogomolnyi:1976kr,Prasad:1975kr,Manton:1977er}. There the forces between monopoles are influenced not only by their structure, but also by the strength of the Higgs field.

The minimum of the potential energy in Eq.~(\ref{Lagr4D}) is at the equatorial $S^2$ of $S^3\cong\,$SU(2). Solitonic solutions are therefore characterised by two topological quantum number counting the number of coverings of $S^2$ and $S^3$. They allow us to classify stable and static solitons.

With the hedgehog ansatz
\begin{equation}\label{RegularIgel}
\vec n(x):=\frac{\vec x}{r},\quad\vec q(x):=\vec n(x)\sin\alpha(x),\quad
q_0:=\cos\alpha(x)
\end{equation}
we can solve the minimisation procedure for a static soliton which leads to the non-linear differential equation
\begin{equation}\label{nlDE}
\frac{(1-\cos^2\alpha)\cos\alpha}{\rho^2}+\partial^2_\rho\cos\alpha
-3\rho^2\cos^5\alpha=0.
\end{equation}

There are four homotopy classes of configurations. In every class there is one solution, respecting the hedgehog ansatz
\begin{equation}\label{exemplare}
q_0=\pm\frac{r_0}{\sqrt{r^2+r_0^2}},\quad
q_i=\pm\frac{x_i}{\sqrt{r^2+r_0^2}}\quad\textrm{with}\quad r^2:=\vec x^{\,2}.
\end{equation}
The vectors $\vec q$ of these configurations in a plane through the origin are depicted in four diagrams in Table~\ref{Tabklassen}. There, $\vec q$-vectors are drawn in full-red for $q_0>0$, dashed-green for $q_0<0$ and approaching black for $q_0\approx0$. The four classes can be transformed to each other by center transformations $Q(\vec r)\gehtzu{z}-Q(\vec r)$ and parity transformations $Q(\vec r)\gehtzu{\Pi}Q(-\vec r)$. The configurations within every class differ by global rotations and/or translations.
\begin{table}[h]\begin{center}\hspace*{-10mm}\begin{tabular}{cccc}\hline 
$\mathcal T=1$&$\mathcal T=z\Pi$&$\mathcal T=z$&$\mathcal T=\Pi$\\\hline
$Z=1$&$Z=1$&$Z=-1$&$Z=-1$\\
$\mathcal Q=\frac{1}{2}$&$\mathcal Q=-\frac{1}{2}$&$\mathcal Q=\frac{1}{2}$&$\mathcal Q=-\frac{1}{2}$\\\hline
\includegraphics[scale=0.25]{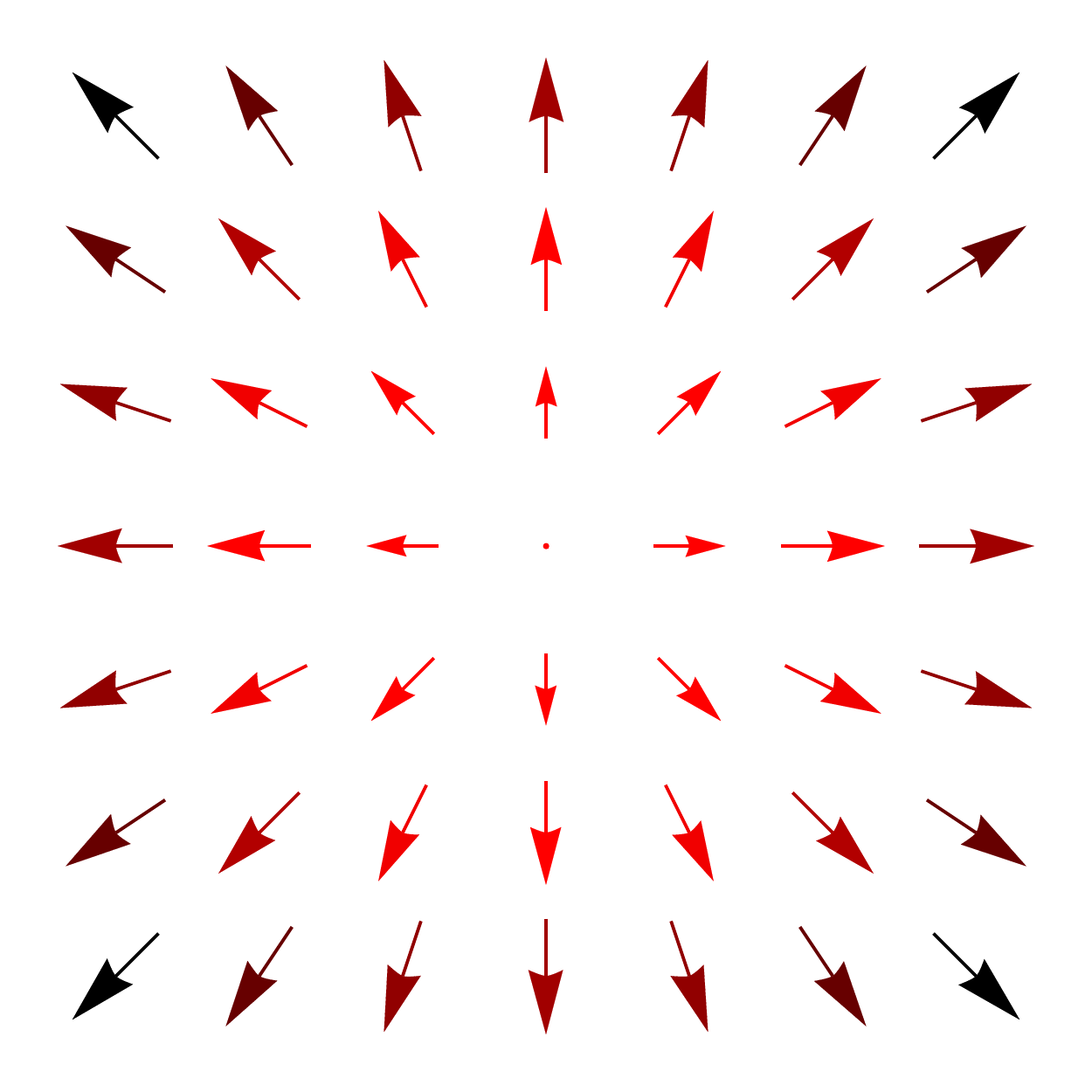}&\includegraphics[scale=0.25]{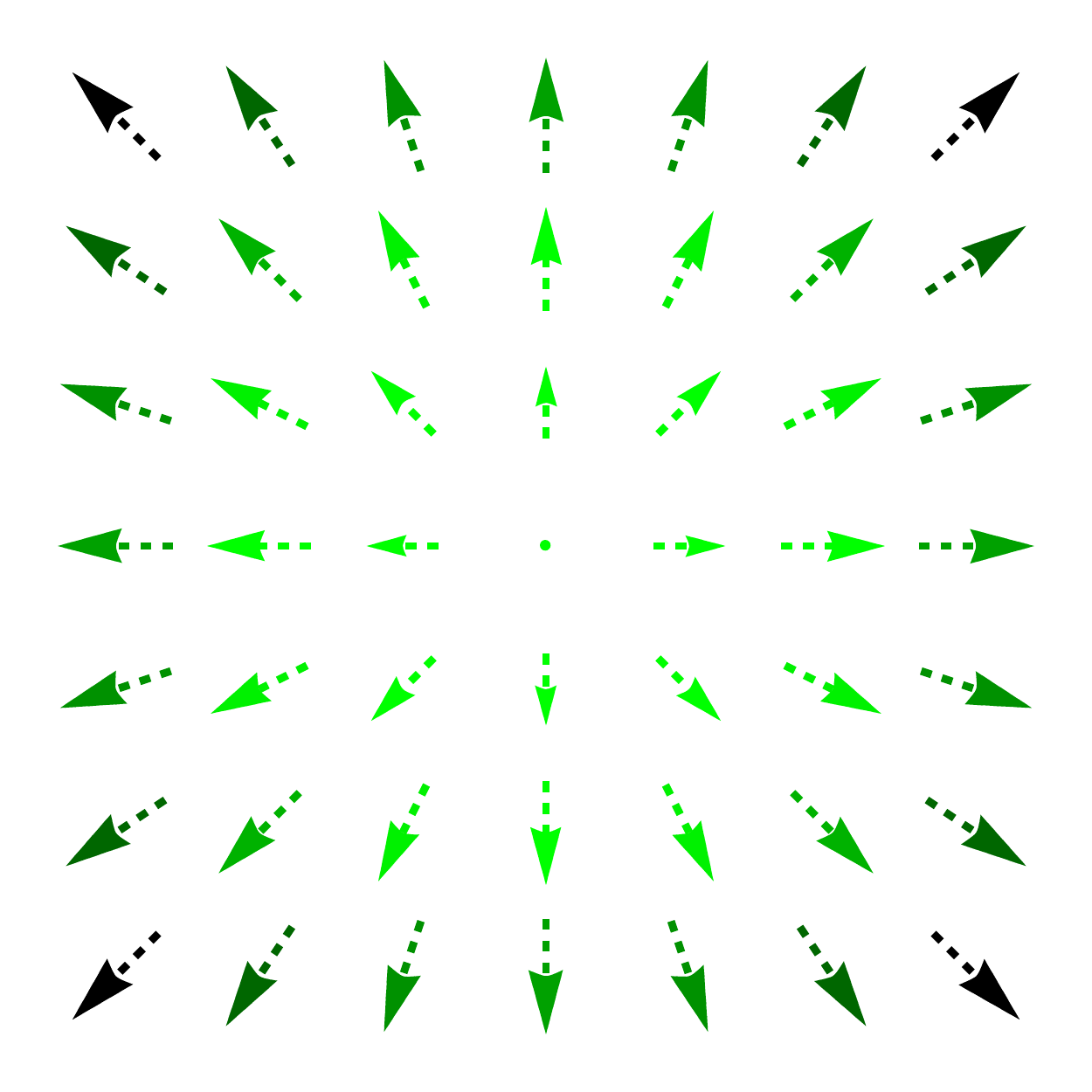}&\includegraphics[scale=0.25]{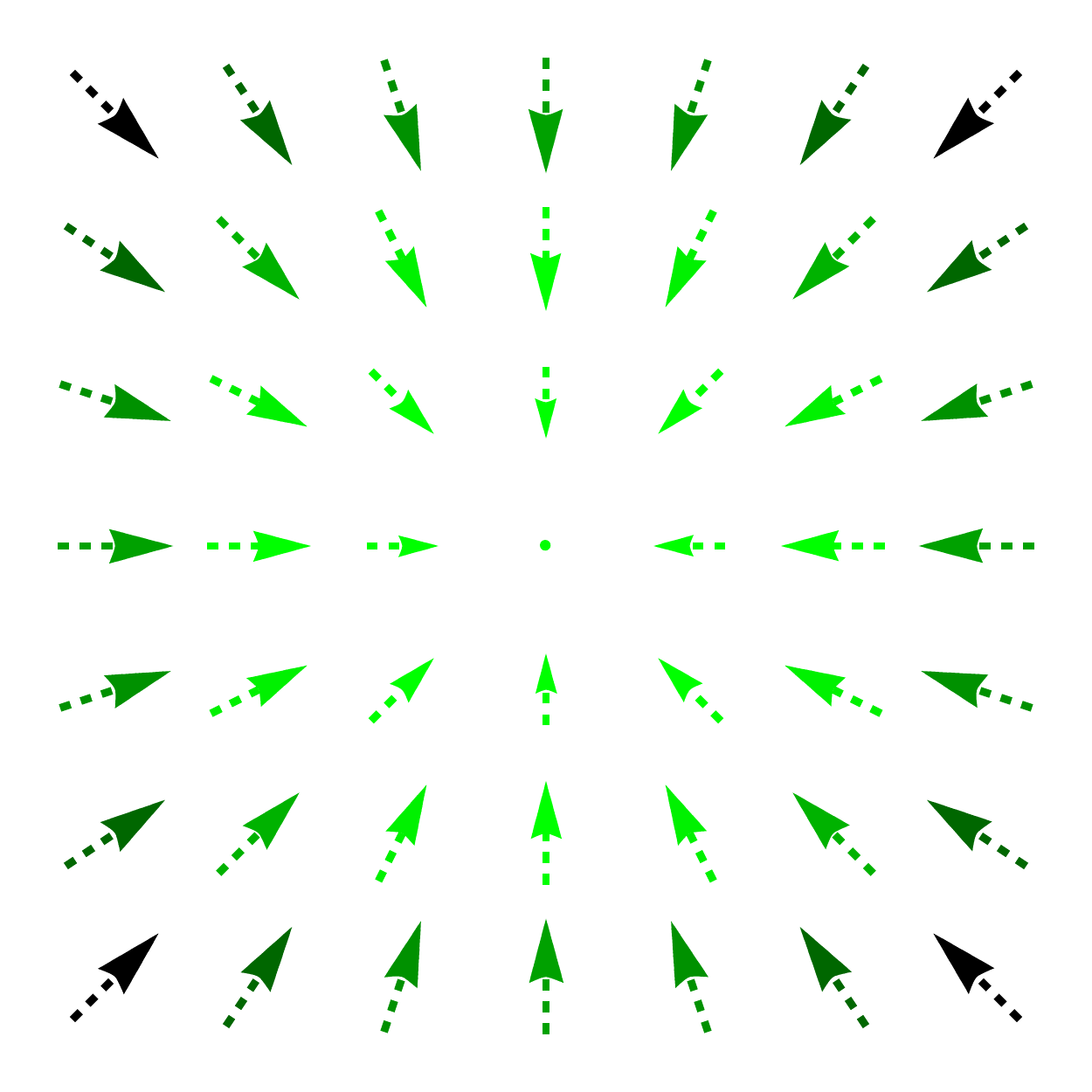}&\includegraphics[scale=0.25]{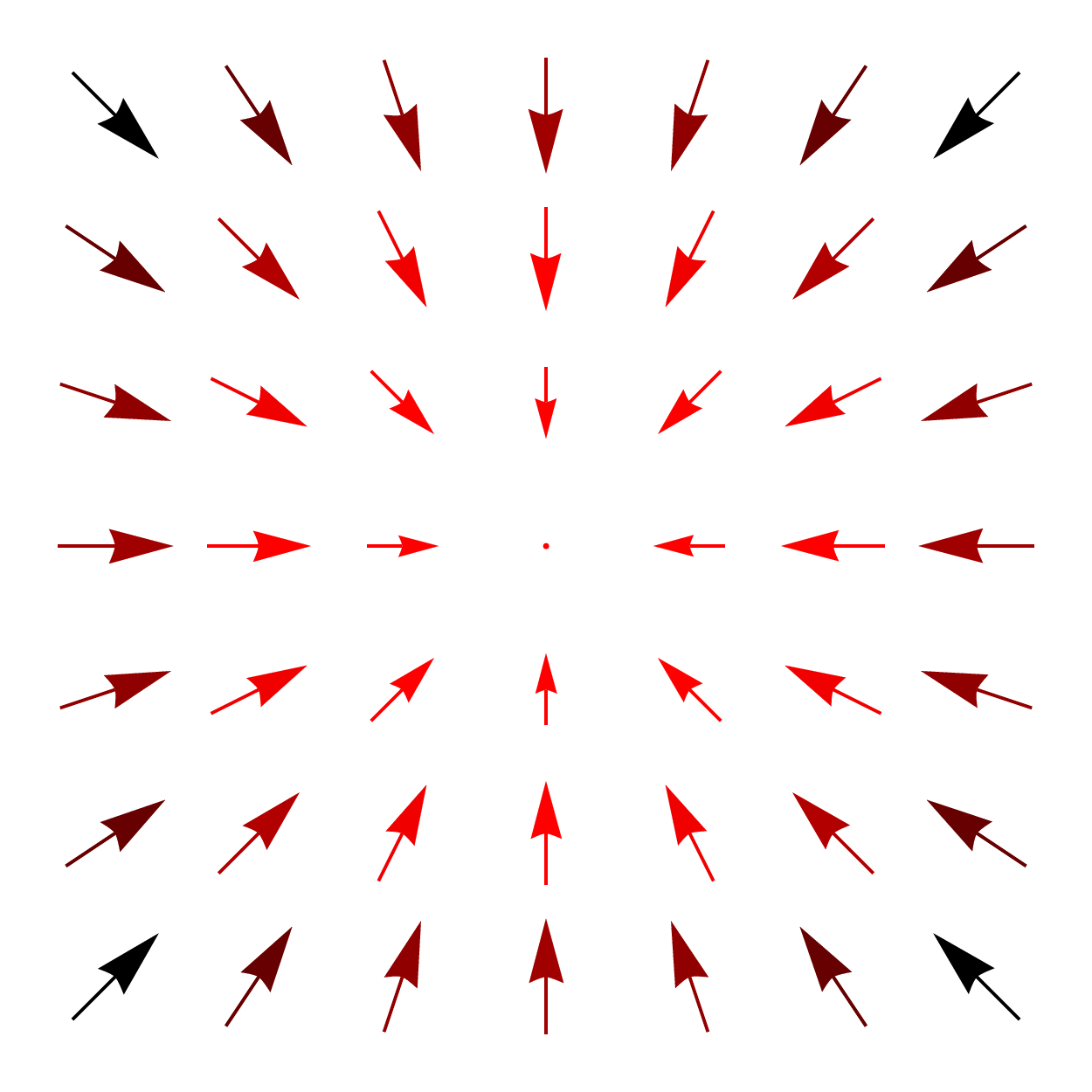}\\\hline
\end{tabular}\caption{The diagrams show the imaginary components $\vec q$ of the field of single solitons, in full red for the hemisphere with $q_0>0$ and in dashed green for $q_0<0$ and all approaching black for $q_0\to0$. The transformations $\mathcal T$ of the hedge-hog configuration in the first column, modify the fields $\vec q$ and $q_0$ and the topological quantum numbers $Z$ and $\mathcal Q$.}
\label{Tabklassen}\end{center}\end{table}
The four classes of solutions are distinguished in the second row of Table~\ref{Tabklassen} by the number $Z$ of electron charges $-e_0$ and in the third row by the chirality $\chi$, the sign of the topological charge $\mathcal Q$. These four soliton types have identical mass~\cite{Faber:1999ia}
\begin{equation}\label{EneMonopFormel}
m_0=\frac{\alpha_f\hbar c_0}{r_0}\frac{\pi}{4}.
\end{equation}
It can be adjusted to the mass $m_e$ of electrons and positrons by choosing
\begin{equation}\label{r0Monop}
r_0=\alpha_f\frac{\hbar}{m_ec_0}\frac{\pi}{4}=~2.21~\textrm{fm}.
\end{equation}
Thus, we can describe the four components of Dirac spinors, electrons $e_-$ and positrons $e_+$, spinup $\uparrow$ and down $\downarrow$. According to the SO(3) description, single isolated soliton configurations differing by a non-trivial center transformation $z$ are identical. In multiple soliton configurations the SU(2)-configuration depends on the other solitons.

It is nice to see that the four classes of solutions agree with the well-known components of Dirac spinors in the standard representation:$e_-^\uparrow, e_-^\downarrow, e_+^\uparrow, e_+^\downarrow$ . The four configurations differ in center transformations $Q\gehtzu{z}-Q$ and in parity transformations $\vec n\gehtzu{\Pi}-\vec n$. In table~\ref{Tabklassen} we observe that a parity transformation corresponds to the product of charge transformation and spin reversal.

Now we would like to clarify which assumptions are necessary, to get the Dirac equation for solitons.

\section{Free Dirac equation}\label{SectFree}
Within relativistic quantum theory the Dirac equation is the basic equation for the understanding of the properties of spin-1/2 fermions. It is also the basis for their understanding within quantum field theory. The solutions $\psi$ of the Dirac equations have the form of spinors consisting of four complex functions for spin-up and spin-down components of fermions and anti-fermions. The outstanding result of the Dirac equation was the explanation of the fine-structure of the hydrogen spectrum and its degeneracy~\cite{akhiezer1965quantum}. A result, later improved only by QED~\cite{PhysRev.72.339}.

In relativistic quantum mechanics we can attribute to particles an intrinsic parity and use $\gamma_0=\textrm{diag}(1,1,-1,-1)$ as parity operator. The Dirac equation has two pairs of solutions of opposite parity. This abstract parity property has a figurative realisation in the diagrams of Tab.~1 with two pairs of configurations which differ by an application of the parity operator $\Pi$.

According to quantum mechanics we assign to the four configurations in Table~\ref{Tabklassen} unit vectors in a four component Hilbert space, and arrange them in four-component spinors $\psi_0$ which fulfil the equation
\begin{equation}\label{parit}
H_0\psi_0:=m_0c_0^2\,\gamma_0\psi_0,
\end{equation}
for the Hamiltonian $H_0$, if we attribute to positrons a negative mass.

Starting from the properties of solitons we will now proceed in detail to a derivation of the Dirac equation for moving solitons. From our knowledge about gamma matrices, the Lorentz transformation of momentum four-vectors and of Dirac spinors, we know the result which we expect. But it may be interesting to see in detail how we can derive the Dirac equation for free particles from the relativistic properties of the soliton four momentum only. In the next section we will then proceed to the interacting case.

\begin{figure}[h!]
\centering
\includegraphics[scale=0.8]{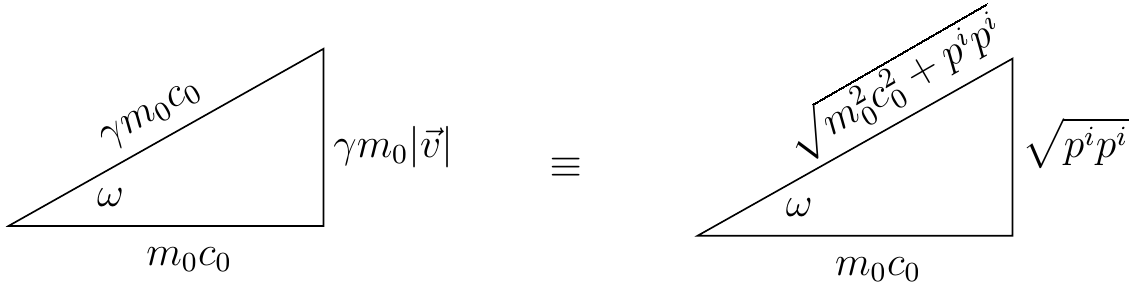}
\caption{Graphical representation of the relations~(\ref{viererImp}) between the four-momentum components and the invariant mass.}
\label{rechtwink}
\end{figure}

As discussed in Ref.~\cite{Faber:1999ia} the mass of solitons moving with classical momentum $\vec p$ increases with the well-known factor $\gamma(\vec p):=1/\sqrt{1-\vec p^2/(m_0c_0)^2}$, as one can understand from the invariance of the MTP-Lagrangian under Lorentz transformations. Thus, we generalise the Hamiltonian in Eq.~(\ref{parit}) to
\begin{equation}\label{bewegt}
H:=\gamma(\vec p)\,m_0c_0^2\,\gamma_0.
\end{equation}
The relativistic four-momentum for a soliton of mass $m_0$ and velocity $\vec v$ reads
\begin{equation}\label{viererImp}
p^\mu=\gamma m_0(c_0,\vec v)^\mu,\quad\gamma=\frac{1}{\sqrt{1-\frac{\vec v^2}{c_0^2}}}.
\end{equation}
These relations can be represented graphically by the right-angled triangles in Fig.~\ref{rechtwink}, resembling the decomposition of a vector into orthogonal components.

Since MTP uses the degrees of freedom of an SO(3)-field and simplifies the calculations by formulations in the defining (two dimensional) representation of SU(2), the MTP-Lagrangian $(\ref{Lagr4D})$ for free particles cannot distinguish between solitons with the same $\mathcal Q$ and different $Z$ which differ by a center transformation only, see Table~\ref{Tabklassen}. Due to this indistinguishability we can apply a unitary transformation between the first and the second pair of solitons, between the upper and the lower component of the Dirac spinor. Since the matrix 
\begin{eqnarray}\label{iDef}
i:=\begin{pmatrix}0&-\mathbbm 1_2\\\mathbbm 1_2&0\end{pmatrix},
\end{eqnarray}
also used for the definition of the symplectic group, acts like an imaginary unit, $i^2=-\mathbbm 1_4$, we can formulate such a "rotation'' with the rotational vector $\vec\omega:=\omega\,\vec e_\omega$ as
\begin{equation}\label{KompTrafo}
U_\omega:=\exp\{i\,\vec\omega\,\frac{\vec\sigma}{2}\}=\cos\frac{\omega}{2}
+\underbrace{\begin{pmatrix}0&-\mathbbm 1_2\\\mathbbm 1_2&0\end{pmatrix}
\vec\sigma}_{-\vec\gamma}\vec e_\omega\,\sin\frac{\omega}{2},
\end{equation}
where the well-known $\vec\gamma$-matrices in the Dirac (or standard) representation appear. It turns out that $U_\omega$ is the expected Lorentz transformation of Dirac spinors, a reversed version of the Foldy-Wouthuysen transformation~\cite{PhysRev.78.29}.

Due to the anticommutation relations between $\gamma^0$ and $\gamma^i$ which appeared in the underbrace of Eq.~(\ref{KompTrafo})
\begin{equation}\label{Antikomm}
\gamma^0\gamma^i+\gamma^i\gamma^0=0\quad\textrm{with}\quad
\gamma^0:=\begin{pmatrix}\mathbbm 1_2&0\\0&-\mathbbm 1_2\end{pmatrix},\quad
\vec\gamma=\begin{pmatrix}0&\vec\sigma\\-\vec\sigma&0\end{pmatrix}
\end{equation}
we get
\begin{equation}\label{trafoU}
U_\omega\gamma^0\iist{KompTrafo}{Antikomm}\gamma^0U_\omega^\dagger.
\end{equation}
The transformation of the Hamiltonian~(\ref{bewegt}) reads therefore
\begin{equation}\label{trafoH}
H\ist{bewegt}\gamma(\vec p)\,m_0c_0^2\,\gamma_0\;\gehtzu{U_\omega}\;
H_\omega:=U_\omega HU_\omega^\dagger\ist{trafoU}H(U_\omega^\dagger)^2.
\end{equation}
with
\begin{equation}\label{rotH}
H_\omega\iist{trafoH}{KompTrafo}\gamma(\vec p)\,m_0c_0^2\,\gamma_0
\exp\{-i\,\vec\omega\,\vec\sigma\}
\iist{KompTrafo}{Antikomm}\gamma(\vec p)\,m_0c_0^2\,\gamma_0
(\cos\omega+\vec\gamma\,\vec e_\omega\sin\omega).
\end{equation}
Since the prefactor of the last bracket is proportional to the hypothenuse $p_0$ of the triangle in Fig.~\ref{rechtwink}, there is a natural choice of $\omega$ and $\vec e_\omega$ given by that triangle. We read from Fig.~\ref{rechtwink} $\cos\omega=1/\gamma(\vec p)$ and $\sin\omega=|\vec v|/c_0$ leading to 
\begin{equation}\label{p0}
p_0:=H_\omega/c_0\ist{rotH}m_0c_0\,\gamma_0+\gamma_0\gamma^ip^i.
\end{equation}
We multiply with $\gamma_0$ from the left, reorder the terms and get
\begin{equation}\label{diracAbgeleitet}
\gamma^\mu p_\mu\ist{p0}m_0c_0.
\end{equation}
From Eq.~(\ref{bewegt}) we conclude that the rotated spinor
\begin{equation}\label{rotpsi}
\psi:=U_\omega\psi_0
\end{equation}
obeys the Dirac equation
\begin{equation}\label{freieDirac}
\gamma^\mu p_\mu\,\psi\iist{diracAbgeleitet}{rotpsi}m_0c_0\,\psi.
\end{equation}

With the usual quantum mechanical laws which also solitons have to obey, coordinates and momenta as canonical conjugate variables, $p_\mu=\mathrm i\hbar\partial_\mu$, and Borns rule, we get the action of free Dirac particles
\begin{equation}\label{freieWirkung}
S_f:=\int \mathrm d^4x\,\bar\psi(\gamma^\mu p_\mu-m_0c_0)\,\psi
\end{equation}
with the free monopole currents
\begin{equation}\label{freieStroeme}
j^\mu:=c_0\bar\psi\gamma^\mu\psi.
\end{equation}

We observe that we get the correct Dirac equation only by choosing $\omega$ as given by the triangle in Fig.~\ref{rechtwink}, as was emphasised in Ref.~\cite{Costella1995}. The correct non-relativistic limit is only retained if the scalar product $\vec\sigma\vec p$ is not destroyed by the transformation~(\ref{KompTrafo}).

Further, we would like to mention, that the solutions of the Dirac equations are the colums of the matrix
\begin{equation}\label{Loesungen}
\gamma^\mu p_\mu+m_0c_0.
\end{equation}

After getting the Dirac equation for free solitons we will now concentrate on their interaction with electromagnetic fields. This will be done mainly in two steps. First, we have to separate the dofs of particles and fields and then we have to make a transition to dual variables.

\section{Interaction with electromagnetic fields}\label{SectInteract}
The MTP-Lagrangian $(\ref{Lagr4D})$ describes multiparticle systems and their fields $f_{\mu\nu}$ on the same footing. To get the Dirac equation considering single point-like charges moving in electromagnetic fields, we have to separate artificially the dofs of charges and fields. This is possible only in the limit of point-like solitons (dual Dirac-monopoles), see Refs.~\cite{Faber:1999ia},\cite{faber:2002nw}, where the soliton field $Q(x)$ is purely imaginary. In this limit we arrive at the formulation of dual Dirac monopoles in the Wu-Yang representation~\cite{wu:1976qk}
\begin{equation}\label{Qlimit}
Q(x)\quad\to\quad Q(x)\ist{Grundfeld}-\mathrm i\vec\sigma\vec n(x),\quad\vec n^2:=1,\quad\vec q\ist{Grundfeld}\vec n
\end{equation}
where the $\vec n$-field takes values on $\mathrm{S}^2$. Thus, the Jacobian of the map from $\mathrm{M}^4\to\mathrm{S}^2$ is given by the curvature field
\begin{equation}\label{CurvWu}
\vec R_{\mu\nu}(x)\quad\to\quad\vec R_{\mu\nu}(x)\iist{RSU2}{Qlimit}
\partial_\mu\vec n(x)\times\partial_\nu\vec n(x).
\end{equation}
which we relate by~\cite{Faber:2022zwv}
\begin{eqnarray}\label{AbelianFS}
{\hspace{1mm}^\star}\hspace{-1mm}f_{\mu\nu}&
=:&-\frac{e_0}{4\pi\varepsilon_0 c_0}\vec R_{\mu\nu}\vec n
\ist{CurvWu}-\frac{e_0}{4\pi\varepsilon_0 c_0}\,\vec n
(\partial_\mu\vec n\times\partial_\nu\vec n).
\end{eqnarray}
to the dual~\footnote{The $^*$-operator in Eq.~(\ref{AbelianFS}) performs a Hodge duality transformation, transforming $\mathbf E/c_0\to\mathbf B$ and $\mathbf B\to-\mathbf E/c_0$ and can be formulated with the four-dimensional $\epsilon$-tensor, compare the underbrace in Eq.~(\ref{EdynGrenJ}).} abelian electromagnetic field strength in SI units. The Lagrange density $\mathcal L_\mathrm{MTP}$ reduces to 
\begin{equation}\label{EDLagrangian}\hspace{-4mm}
\mathcal L_\mathrm{MTP}\quad\to\quad \mathcal L_\mathrm{ED}\iist{Lagr4D}{Qlimit}
-\frac{\alpha_f\hbar c_0}{4\pi}\frac{1}{4}\,
(\vec n\vec R_{\mu\nu})(\vec n\vec R^{\mu\nu})
\ist{AbelianFS}-\frac{1}{4\mu_0}{\hspace{0.5mm}^\star}\hspace{-0.5mm}
f_{\mu\nu}(x){\hspace{0.5mm}^\star}\hspace{-0.5mm}f^{\mu\nu}(x).
\end{equation}
$\mathcal L_\mathrm{dual}$ differs in the sign only from the Lagrangian of Maxwell's electrodynamics.

We use the singularities of the $\vec n$-field at given time $t$ to identify the soliton currents. These singularities are located in infinitesimal space-like three dimensional volumes, the centers of hedgehog configurations of the $\vec n$-field. We describe them by $N$ time-like world-lines of singularities
\begin{equation}\label{worldline}
X_i^\mu(\tau)\qquad i=1,\cdots,N,
\end{equation}
evolving in time with velocities smaller than $c_0$  from $t=-\infty$ to $t=+\infty$, or meeting with lines of opposite charge, where particle pairs are created or annihilated. This makes it possible to define for each $\vec n$-field configuration a singular vector current along the above world-lines
\begin{equation}\label{divFluss}
\sum_{i=1}^NZ_i\int\mathrm d\tau\,\frac{\mathrm dX_i^\kappa(\tau)}{\mathrm d\tau}
\,\,\delta^4(x-X_i(\tau)):=\frac{1}{8\pi}\varepsilon^{\kappa\lambda\mu\nu}\,
\partial_\lambda\{\vec n(\partial_\mu\vec n\times\partial_\nu\vec n)\}.
\end{equation}
Integrals over arbitrary spatial volumes $V$ over the left and the right side of this equation equate the number of world lines crossing the volume with the number of coverings of $\mathrm S^2$ by the $\vec n$-field at the volumes surface $\partial V$. For a single soliton resting at the origin with $Z=1,\tau=t,X^\kappa(\tau)=c_0t\,\delta_0^\kappa$ both sides reduce to a three-dimensional delta-function at the origin $\delta^3(\vec x)\delta_0^\kappa$ for the time-component of the singular vector current. Multiplying the world lines of delta-functions with $-e_0c_0$ we define the electric current density
\begin{equation}\label{Defj}
q\,j^\kappa:=-e_0c_0\sum_{i=1}^NZ_i\int\mathrm d\tau\,
\frac{\mathrm dX_i^\kappa(\tau)}{\mathrm d\tau}\,\,\delta^4(x-X_i(\tau))
\end{equation}
of point-like charges $q_i=-e_0Z_i$. A further multiplication by $\mu_0$ leads to
\begin{equation}\label{EdynGrenJ}
\mu_0\,q\,j^\kappa\ist{divFluss}-\partial_\lambda\underbrace{\frac{e_0 c_0\mu_0}{4\pi}
\frac{1}{2}\epsilon^{\kappa\lambda\mu\nu}\,
\vec n[\partial_\mu\vec n\times\partial_\nu\vec n]}_{f^{\kappa\lambda}}
\ist{AbelianFS}\partial_\lambda f^{\lambda\kappa}.
\end{equation}
The contributions of these inhomogeneous Maxwell equations are non-zero along the world-lines~(\ref{worldline}) only and absorb all currents in the divergencies of $f^{\mu\nu}$. After the transition~(\ref{Qlimit}) from extended monopoles to point-like particles, the Hodge transformation of Eq.~(\ref{Qlimit}) respecting  ${\hspace{0.5mm}^{\star\star}}\hspace{-0.5mm}f^{\mu\nu}=-f^{\mu\nu}$ has simplified Eqs.~(\ref{EdynGrenJ}). To keep the structure $L=T-V$ of Lagrangians we flip the sign of the Lagrangian $\mathcal L_\mathrm{ED}$ which after the Hodge transformation reads
\begin{equation}\label{TeileLagrangian}
\mathcal L:=-\mathcal L_\mathrm{ED}\ist{EDLagrangian}
-\frac{1}{4\mu_0}\,f_{\mu\nu}f^{\mu\nu}.
\end{equation}

For the next step, the introduction of the electro-magnetic four-potential $A_\mu$, we need the homogeneous Maxwell equations. This deserves a further argument. The idea behind MTP is that particles are topological solitons and characterised by topological quantum numbers. Topological stable objects can not be removed from the configurations by the minimisation of the energy. They can be removed by the fusion with appropriate antiparticles only. In MTP, the electric charge is such a topological property related to the map $\Pi_2(\mathrm S^2)$, see Eq.~(\ref{divFluss}). In the limit~(\ref{Qlimit}) MTP has two degrees of freedom only, compared to 4 degrees of freedom of $A_\mu$. This reduces the flexibility to fulfil the abelian Bianchi identity and may lead to non-vanishing conserved magnetic currents
\begin{equation}\label{magcur}
g^\mu:=c\,\partial_\nu \hspace{0.2mm}{^*}\hspace{-0.2mm}f^{\nu \mu}\quad
\textrm{with}\quad\partial_\mu g^\mu=0.
\end{equation}
Since they are not stabilised by topological quantum numbers their contributions are minimised by the least action principle and the equations of motion, see Refs.~\cite{Faber:1999ia,faber:2002nw},
\begin{equation}\label{EOM}
\partial_\mu\vec{n}\;g^\mu=0\quad\Rightarrow\quad
\hspace{0.8mm}{^*}\hspace{-0.8mm}f_{\mu \nu} g^\nu=0.
\end{equation}
The second set of equations proves that the dual Coulomb and Lorentz forces on magnetic currents are vanishing. We show now that the magnetic currents do not contribute to Coulomb and Lorentz forces on electric currents and thus can be neglected in the formulation of their dynamics.

We start from the observation that the divergence of the total energy momentum tensor~(\ref{DefEMT}) is vanishing. After the separation of dofs of charges and their fields the divergence of the energy momentum tensor of the fields
\begin{equation}\label{ExpEMomTen}
\Theta^\mu_{\;\nu}(x)\to T^\mu_{\;\nu}(x):=
-\frac{1}{\mu_0}{\hspace{0.5mm}^\star}\hspace{-0.5mm}f_{\nu\sigma}(x)
 {\hspace{0.5mm}^\star}\hspace{-0.5mm}f^{\mu\sigma}(x)-\mathcal L_{\rm ED}(x)\,\delta^\mu_\nu,
\end{equation}
leads therefore as a reaction to the force density on charges
\begin{equation}\begin{aligned}\label{forcedensityderivation}
f^\mu_\mathrm{e}&:=-\partial^\nu T^\mu_{\;\nu}\ist{ExpEMomTen}
\frac{1}{\mu_0}\partial^\nu\left({\hspace{0.5mm}^\star}\hspace{-0.5mm}
f_{\nu\rho}{\hspace{0.5mm}^\star}\hspace{-0.5mm}f^{\mu\rho}\right)
-\frac{1}{4\mu_0}\partial^\mu({\hspace{0.5mm}^\star}\hspace{-0.5mm} f_{\rho\nu}
{\hspace{0.5mm}^\star}\hspace{-0.5mm}f^{\rho\nu})=\\
&=\frac{1}{\mu_0}
[\underbrace{\partial^\nu{\hspace{0.5mm}^\star}\hspace{-0.5mm}f_{\nu\rho }}
_{\frac{1}{c_0} g_\rho}{\hspace{0.5mm}^\star}\hspace{-0.5mm}f^{\mu\rho}
+\underbrace{{\hspace{0.5mm}^\star}\hspace{-0.5mm}f_{\nu\rho}
\;\partial^\nu{\hspace{0.5mm}^\star}\hspace{-0.5mm}f^{\mu\rho}}_{
-{\hspace{0.5mm}^\star}\hspace{-0.5mm}f_{\nu\rho}
\partial^\rho{\hspace{0.5mm}^\star}\hspace{-0.5mm}f^{\mu\nu}}
+\frac{1}{2}{\hspace{0.5mm}^\star}\hspace{-0.5mm}f_{\nu\rho}
\;\partial^\mu{\hspace{0.5mm}^\star}\hspace{-0.5mm}f^{\rho\nu}]=\\
&\ist{magcur}\frac{1}{\mu_0 c_0}\underbrace{
{\hspace{0.5mm}^\star}\hspace{-0.5mm}f^{\mu\rho} g_\rho}_{0}
+\frac{1}{2\mu_0}{\hspace{0.5mm}^\star}\hspace{-0.5mm}f_{\nu\rho}
[\underbrace{\partial^\nu{\hspace{0.5mm}^\star}\hspace{-0.5mm}f^{\mu\rho}
+\partial^\rho{\hspace{0.5mm}^\star}\hspace{-0.5mm}f^{\nu\mu}
+\partial^\mu{\hspace{0.5mm}^\star}\hspace{-0.5mm}f^{\rho\nu}}_{
-\mu_0\,q\,\epsilon^{\mu\nu\rho\sigma} j_\sigma}]=\\
&\iist{EOM}{EdynGrenJ}-\frac{q}{2}\epsilon^{\mu\nu\rho\sigma}
{\hspace{0.5mm}^\star}\hspace{-0.5mm}f_{\nu\rho} j_\sigma 
=f^{\mu\sigma} j_\sigma q.
\end{aligned}\end{equation}
Electric and magnetic fields cause therefore the only forces on electric currents. According to these arguments we can neglect magnetic currents in the description of the dynamics of electric currents and assume the validity of the homogeneous Maxwell equations by formulating
\begin{equation}\label{abelscheDef}
f_{\mu\nu}:=\partial_\mu a_\nu-\partial_\nu a_\mu.
\end{equation}

To complete the separation of dofs of charges and fields we have to divide $f_{\mu\nu}$ in internal contributions $\mathcal F_{\mu\nu}$ related to the soliton currents $j^\mu$ and external fields $F_{\mu\nu}$
\begin{equation}\label{composition}
f_{\mu\nu}:=\mathcal F_{\mu\nu}+F_{\mu\nu},\quad a_\mu:=\mathcal A_\mu+A_\mu
\end{equation}
For the Lagrangian, this separation leads to three contributions, the action density of external electromagnetic fields ($F^2$), the interaction term between char\-ges and fields ($\mathcal F\,F$) and the Lagrange density~(\ref{freieWirkung}) of free particles ($\mathcal F^2$). Since the internal contribution $\mathcal F_{\mu\nu}$ exhausts the field of all point-like sources, the corresponding infinite self-energies have to be absorbed in the Lagrangian of the free particles and substituted by the finite experimental mass $m_0$. This results in the equivalence, derived in Eq.~(\ref{freieWirkung}),
\begin{equation}\label{freieLagrangian}
-\frac{1}{4\mu_0}\int \mathrm d^4x\mathcal F_{\mu\nu}\mathcal F^{\mu\nu}
\equiv\int \mathrm d^4x\,\bar\psi(\gamma^\mu p_\mu-m_0c_0)\,\psi
\ist{freieWirkung}S_f.
\end{equation}
The interaction term of the action gets therefore
\begin{equation}\begin{aligned}\label{WW}
S_\mathrm{int}:=&-\frac{1}{2\mu_0}\int\mathrm d^4x\,F_{\mu\nu}\mathcal F^{\mu\nu}
\ist{abelscheDef}\frac{1}{\mu_0}\int\mathrm d^4x\,A_\nu\partial_\mu
\mathcal F^{\mu\nu}=\\&\ist{EdynGrenJ}\int\mathrm d^4x\,q\,A_\nu j^\nu
\ist{freieStroeme}\int\mathrm d^4x\,q\,A_\nu\bar\psi\gamma^\nu\psi.
\end{aligned}\end{equation}

Collecting the three action contributions we get from the Lagrangian~(\ref{TeileLagrangian}) the total action for a Dirac particle in an external vector field $A_\mu$
\begin{equation}\label{Sgesamt}
S\iist{freieLagrangian}{WW}
\int\mathrm d^4x\,\Big\{\bar\psi\big[\gamma^\mu(p_\mu-qA_\mu)-m_0c_0\big]\psi
-\frac{1}{4\mu_0}\,F_{\mu\nu}F^{\mu\nu}\Big\}.
\end{equation}

\section{Conclusion}
The presently accepted theory, the Standard Model of Particle Physics, one can see from two vantage points
\begin{itemize}
\item it provides excellent fits to scattering data,
\item it embrassingly predicts the vacuum energy density of the cosmos
  off by a factor of $10^{50}$--$10^{120}$ \cite{Martin_2012} and has
  nothing to say about the classical selfenergy of its fundamental massive fermions.
\end{itemize}

The idea of MTP is to arrive at a
geometric formulation of particle physics. If successful this would
\begin{itemize}
\item provide a geometrical basis for the standard model,
\item move particle physics closer to gravity and facilitate unification,
\item show in detail that particles and their fields are describable by
   the same degrees of freedom,
\item explain why algebras are so important in particle physics. They allow to
   work in the tangential spaces of the corresponding group manifolds
   supplying the deeper degrees of freedom for the common description of
   particles and their fields,
\item explain gauge symmetries as basis changes in the manifolds of the
   corresponding groups,
\item allow to understand the origin of cosmological phenomena like
   inflation, see Sect.~6.4 of Ref.~\cite{Faber:2022zwv},
\item suggest a solution for the cosmological constant problem,
  substituting the constant by a function, the potential energy density of
  solitons,
   see Sect.6.5 of Ref.~\cite{Faber:2022zwv},
\item allow to understand that dark matter is not consisting of particles
   but of non-quantised lumps of matter, e.g. alpha waves in MTP as
   suggested in Sect. 6.3 of Ref.~\cite{Faber:2022zwv}.
\end{itemize}
I have presented the interesting consequences of MTP in
section 7.2 of Ref.~\cite{Faber:2022zwv}.

To summarise, with MTP we can describe electric charges, quantised already at the classical level in units of the elementary charge, by topological solitons of finite size and self-energy, without any divergencies. In this model we get four classes of stable solitons which nicely correspond to the four components of Dirac electrons. This begs for an answer under which conditions and approximations the Dirac equation can be formulated for these solitons. Since the Dirac equation is the basic equation of relativistic quantum mechanics it is obvious that the basic concepts of quantum mechanics have to be respected. Hilbert space vectors have to be attributed to the soliton states, Born's rule has to be applied to quantum mechanical probabilities and the canonical commutation relations to coordinates and momenta. To get the Dirac equation for moving non-interacting particles it turned out to be essential that the MTP Lagrangian can not distinguish between solitons differing by a center transformation only.

In the Dirac equation for interacting particles charges and electromagnetic fields are formulated by different degrees of freedom. This requests for a corresponding separation of degrees of freedom in the MTP. This is achieved in the limit when solitons are approximated by dual point-like Dirac monopoles. The problematic consequences of the singularity of these point-like charges are avoided by attributing to the free fermions the measured masses of electrons. In this limit the difference between electrons and positrons is lost and unitary transformations between them are possible. Electrons and positrons can be distinguished by the sign of the mass term. The interaction between charges and fields is implemented in Schrödinger and Dirac equations by the covariant derivative. It is impossible to obtain this type of interaction from the MTP just by using its dual formulation in the presence of non-vanishing magnetic currents. We show that the dynamics of charges in the MTP is determined by Coulomb and the Lorentz forces only. Since it is not influenced by magnetic currents they can be neglected. The duality transformation allows us then to formulate electromagnetic fields in terms of the usual Abelian vector potentials.

We conclude that in the limit of point-like charges, after absorbing the infinite self-energy into the experimental masses, the interaction between charges is described by a $1/r$ potential. On the other hand the finite size of solitons of the MTP leads to a running of the charge and to a modification of the $1/r^2$ force. This supports the idea to implement the soliton-antisoliton potential energy of the MTP in Schr\"odinger and Dirac equations and to determine in this singularity free description the modifications of the eigenvalues of the stationary equations and to compare them to the experimental spectra, especially to the size of the Lamb shift.

A completely different construction of electrons was suggested in Ref.~\cite{Hofmann:2020yhn} and a different attempt to derive the Dirac equation was undertaken in Ref.~\cite{Close2015}.

\section*{Acknowledgement}
I want to thank to Dmitry Antonov for pointing my attention to the Foldy-Wouthuysen transformation. To the three referees and the scientific editor I am grateful for their strong, but very constructive critics and to Rudolf Golubich for careful reading of the manuscript.

\bibliographystyle{utphys}
\bibliography{diracEq}

\end{document}